\documentclass[aps.prb,twocolumn,floatfix,superscriptaddress,showpacs]{revtex4-1}
\usepackage{amsfonts,amsmath,amssymb}
\usepackage{bm}
\usepackage{pgf}
\usepackage{dcolumn}

\graphicspath{{mpl_plots/}}

\begin{document}
\title{Ordering Behavior of the Two-Dimensional Ising Spin Glass with Long-Range Correlated Disorder}
\author{L. M\"unster}
\affiliation{
Institut f\"ur Physik, Universit\"at Oldenburg, 26111 Oldenburg, Germany
}
\author{C. Norrenbrock}
\affiliation{
Institut f\"ur Physik, Universit\"at Oldenburg, 26111 Oldenburg, Germany
}
\author{A. K. Hartmann}
\email{alexander.hartmann@uni-oldenburg.de}
\affiliation{
Institut f\"ur Physik, Universit\"at Oldenburg, 26111 Oldenburg, Germany
}
\author{A. P. Young}
\affiliation{
University of California Santa Cruz, CA 95064, USA
 }
\date{\today}

\begin{abstract}
The standard two-dimensional Ising spin glass does not exhibit an
ordered phase at finite temperature.  Here, we investigate whether
long-range correlated bonds change this
behavior. The bonds are drawn from a Gaussian distribution with
a two-point correlation for bonds at distance $r$ that decays as
$(1+r^2)^{-a/2}$, $a \geq0$. We study numerically with exact algorithms
the ground state and domain wall
excitations.  Our results indicate that the inclusion of bond correlations
does not lead to a spin-glass
order at any finite temperature. A further analysis reveals that bond
correlations
have a strong effect at local length scales, inducing
ferro/antiferromagnetic domains into the system. The length scale of
ferro/antiferromagnetic order diverges exponentially as the
correlation exponent approaches a critical value, $a \longrightarrow
a_{\text{crit}}=0$. Thus, our results suggest that the system becomes
a ferro/antiferromagnet only in the limit $a\to 0$.
\end{abstract}

\pacs{75.40.Mg, 02.60.Pn, 68.35.Rh}
\maketitle

\section{Introduction}
Spin glasses are disordered magnetic materials which exhibit peculiar
properties at very low temperatures \cite{VincentDupuis2018}.
To understand these materials, the Edwards-Anderson (EA) model and the
Sherrington-Kirkpatrick (SK) model \cite{EdwardsAnderson1975,SherringtonKirkpatrick1975} have been developed. Spin glasses exhibit essential aspects of complex behavior \cite{NewmanStein2013} and research on spin glasses \cite{MezardParisiVirasoro1987,Young1998,BolthausenBovier2007} has stimulated progress in numerous other fields, such as information processing \cite{Nishimori2001}, neuronal networks \cite{Dotsenko1995}, discrete optimization \cite{HartmannRieger2002,HartmannRieger2004} and Monte-Carlo simulation \cite{KatzgraberEtAl2015}.

In this work we study the two-dimensional EA spin glass model with Ising
spins. This model has
short-range quenched random pair-wise interactions, described in detail in Sec.\
\ref{sec:model}.
Its properties are well described in the framework of the scaling/droplet picture \cite{Mcmillan1984,BrayMoore1987,FisherHuse1988},
as has been confirmed by numerical calculations for large
systems using exact ground-state algorithms \cite{MooreHartmann2003Corrections,droplets_long2004}.
The model exhibits no finite-temperature spin glass phas,e in contrast to the three or higher dimensional variants \cite{YoungStinchcombe1976,YoungSouthern1977,HartmannYoung2001}.

At the zero-temperature
phase transition, the distribution of the interaction disorder,
in particular differences between continuous Gaussian and discrete bimodal $\pm J$ disorder
distributions have been the subject
of intensive research
\cite{KhoshbakhtWeigel2018,ParisenEtAl2011,CreightonHuseMiddleton2011,FernandezEtAl2016}.
Since the non-existence of a finite-temperature spin-glass phase for short
range two-dimensional models is independent of the disorder distribution, we
consider here only the Gaussian case. Previous works have shown how the
increase of the mean of the Gaussian from zero to a sufficiently large
value induces a ferromagnetic phase
\cite{SherringtonSouthern1975,rand_bond2004,Edw2007,GarelMonthus2014,gs_perc2020}. 

In this work we address the question how long-range correlations in the interactions
(bonds)
affects the ordering behavior,
in particular whether it leads to a low-temperature spin-glass phase.
Here, long-range means that the bond correlation decays with a power law and
so does not have a
characteristic
length scale. For disordered ferromagnets the effects of long-range
correlations have already been taken studied \cite{WeinribHalperin1983}.

Our study was partially motivated by a corresponding
numerical study 
of the three-dimensional random-field Ising model with
long-range correlation \cite{AhrensHartmann2011},
where, for strong correlation, an influence on the quantitative ordering behavior has been observed
for some critical exponents.
Note, however, that the case of the random-field model is a bit different,
because the correlation acts on the random fields which provide
local randomness competing against long-range ferromagnetic order.
Consequently, from an extended Imry-Ma argument it was predicted
\cite{Nattermann1998} that the random-field Ising model with strong
long-range correlation will show an
increase of the lower critical dimension for
ferromagnetic order. We note that there is no analogous prediction for
the spin-glass case.

The following content is structured into three parts. First, the model is
introduced and it is outlined how ground state computations under changing
boundary conditions are used to produce domain wall excitations. Second, the
results of the simulations will be presented. Finally we give a discussion.

\section{Model and Methods\label{sec:model}}
\subsection{The Ising Spin Glass with Correlated Bonds}
The Ising spin glass consist of Ising spins
$s_{\bm{m}} \in \{ \pm 1\}$ on the sites $\bm{ m} \in \Lambda$
of a two-dimensional lattice, i.e. $\Lambda \subset \mathbb{Z}^2$. 
In this study
only square systems are considered, such that the
spin glass has $L$ spins in each direction and
$\vert \Lambda\vert = L^2$.
The Hamiltonian is given by
\begin{align}
H_{\bm{J}}(\bm{s})=-\sum_{\{ \bm{m},\bm{n}\} \in \mathcal{M} }J_{\bm{m},\bm{n}} s_{\bm{m}}s_{\bm{n}}\,,
\label{eq:hamiltonian}
\end{align}
where the sum runs over all pairs $\mathcal{M}$ of
nearest-neighbor spin sites with periodic boundary conditions (BC) in one
and free boundary conditions in the other direction.
The bonds, $J_{\bm{m},\bm{n}}$, which represent the interaction between two
spins, are random in strength and sign but remain constant over time. Hence,
one speaks of a quenched disorder where the system is investigated under a
fixed realization of the bonds, $\bm{J}$. Here, the bonds originate from a
Gaussian random field, $J(\bm{x})$, which is a function of a
\textit{continuous} position $\bm{x}$, and has zero mean,
$\langle J(\bm{x})\rangle=0$ and a covariance given by

\begin{align}
    \langle J(\bm{x})J(\bm{x}+\bm{r}) \rangle=(1+r^2)^{-a/2},
\end{align}
$\bm{x},\bm{r}\in \mathbb{R}^2$, $a\geq 0$ and $r=\lVert \bm{r}\rVert$. The entries of $\bm{J}$ are given by, $J_{\bm{m},\bm{n}}=J((\bm{m}+\bm{n})/2)$, which ensures that the correlation decays in the same manner along both axes.
The correlation exponent, $a$, is the only parameter to control the
correlation.
For $a=0$ one obtains the Ising model of a ferromagnet or antiferromagnet, respectively, depending on the bond realization. When $a \longrightarrow \infty$ the uncorrelated Ising spin glass model with Gaussian disorder is recovered.
\begin{figure}
    \begin{center}
        \includegraphics{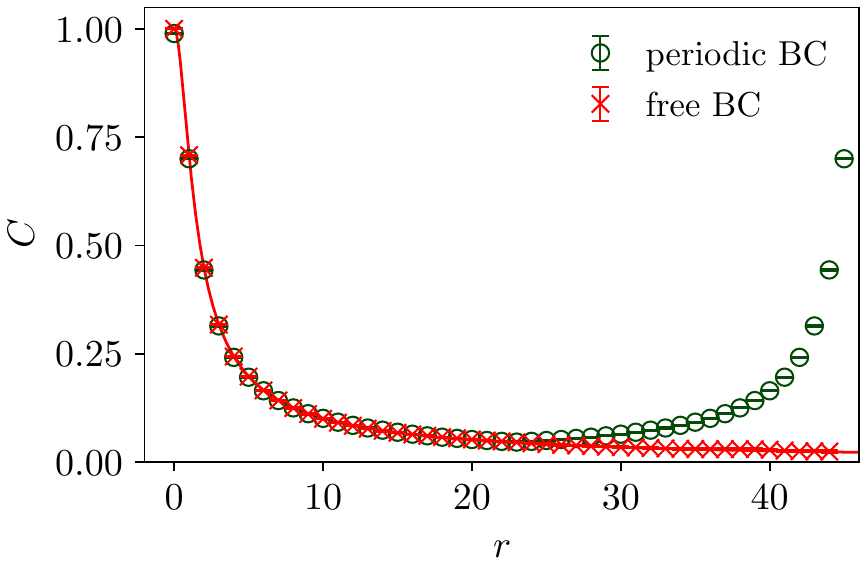}
        \caption{(color online) Correlation of the bonds of a spin glass,
          calculated with Eq. \eqref{eq:bond_corr_estimator}, with
          $a=1$ and $L=46$ spins in each direction, $x$ and $y$.
          The system has free boundary conditions in one direction and periodic boundary conditions in the other direction. The bond correlation was generated according to the
	  Fourier Filtering Method (FFM)
	  \cite{StanleyEtAl1996,AhrensHartmann2011} with periodic boundary conditions in both directions, but
          the size in the directions of free boundaries was chosen much larger
          than the corresponding system size $L$.
          The line follows a fit of type $B_C(1+r^2)^{-A_C/2}$, yielding $B_C=1.0003(12)$ and $A_C=0.9953(13)$. The average was taken over 10000 realization of the disorder. The good agreement proves that the generation of the randomness
        works well.}
        \label{fig:bond_correlation}
    \end{center}
\end{figure}

To generate the correlated bonds numerically we utilized the Fourier Filtering Method (FFM) \cite{StanleyEtAl1996,AhrensHartmann2011}. The FFM is a procedure to create stationary correlated random numbers of previously independent random numbers. Because it is based on the convolution theorem it is possible to benefit from the computationally
efficient Fast Fourier Transform Algorithm \cite{CooleyTukey1965}. For its implementation we relied upon the
functions of the FFTW library, version 3.3.5 \cite{FrigoJohnson2005}. Figure \ref{fig:bond_correlation} shows the average bond correlation along the main axes of a system with $\vert \Lambda\vert= 46^2$ spins calculated by the estimator,
\begin{align}
    C(\bm{r})=\frac{1}{\vert \mathcal{M}'(\bm{r}) \vert} \sum_{\{ \bm{n},\bm{m}\} \in \mathcal{M}'(\bm{r})} \left\langle J_{\bm{m},\bm{n}} J_{\bm{m}+\bm{r},\bm{n}+\bm{r}}  \right\rangle_J~.
    \label{eq:bond_corr_estimator}
\end{align}
Here $\langle...\rangle_J$ denotes the average with respect to the disorder.
$\mathcal{M'}(\bm{r})\subset \mathcal{M}$ contains those bonds
$\{\bm{m},\bm{n}\}$ for which
the bond $\{\bm{m}+\bm{r},\bm{n}+\bm{r}\}$ is also on the lattice. The fact
that
$\mathcal{M'}(\bm{r})$ does not contain all the
bonds $\{ \bm{n},\bm{m}\}$ in $\mathcal{M}$ is due
to the free boundary conditions in one direction. The point is that for vectors $\bm{r}$
which are not exactly parallel to the free boundary, the bond 
$\{\bm{m}+\bm{r},\bm{n}+\bm{r}\}$
does not exist for
all bonds of $\mathcal{M}$. 
For the correlations shown in
Figure \ref{fig:bond_correlation} we only investigated the directions
parallel and perpendicular to the free boundary.

\subsection{Ground States and Domain Walls}
The nature of the ground state (GS) of the two-dimensional Ising spin glass is
an intriguing subject on its own \cite{NewmannStein2000,ArguinDamron2014}.
Furthermore, GS computations of finite systems are a well established tool
\cite{HartmannRieger2002,HartmannRieger2004} to investigate the glassy
behavior of the model in the zero-temperature limit \cite{Hartmann2008}. The
GS is the spin configuration which minimizes Eq. \eqref{eq:hamiltonian} for a
given realization of the bonds. In case of two-dimensional planar lattices
there exist exact procedures to generate the GS with a polynomial worst-case
running time. This is in contrast to the three or higher-dimensional variants
which belong to the class of NP-hard problems \cite{Barahona1982}. 

There is more than one approach to compute the GS of two-dimensional
planar spin glasses,
such as the algorithm of Bieche \emph{et~al.} \cite{BiecheEtAl1980} and that of
Barahona \emph{et~al.} \cite{BarahonaEtA1982}. The key idea of these algorithms is to create a mapping from the original problem defined on the underlying lattice graph of the spin glass
onto a related graph which is constructed in such a manner that the GS can be extracted from a \emph{minimum-weight perfect matching},
which is polynomially computable. In this work we applied an ansatz which includes Kasteleyn-city subgraphs into the mapping process and thus is more efficient \cite{PardellaLiers2008,ThomasMiddleton2007}
in terms of speed and memory usage than the above mentioned algorithms.
This allowed us to investigate systems up to a linear system size of $L=724$
spins in each direction without needing excessive computational resources.
For the computation of the minimum-weight perfect matching the Blossom
$\MakeUppercase{\romannumeral 4}$ algorithm \cite{CookRohe1999} implemented by
A. Rohe \cite{ConcordeBlossom} was utilized.

\begin{figure}[h!t]
    \begin{center}
        {\large $a\longrightarrow \infty$}
        \includegraphics{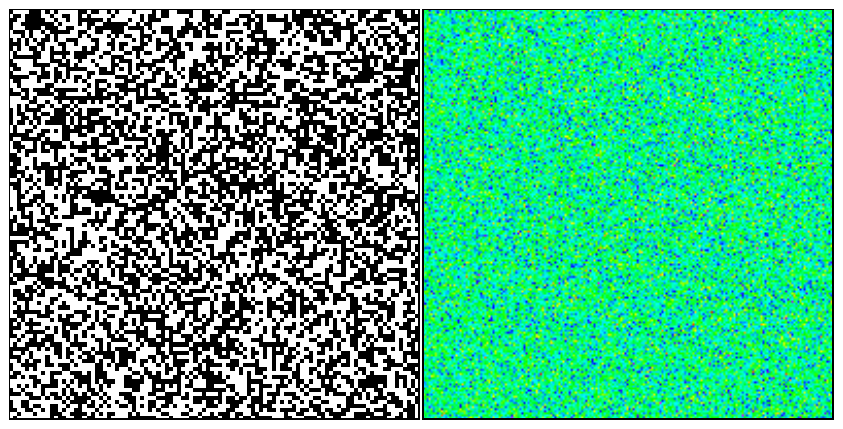}
        {\large $a=1$}
        \includegraphics{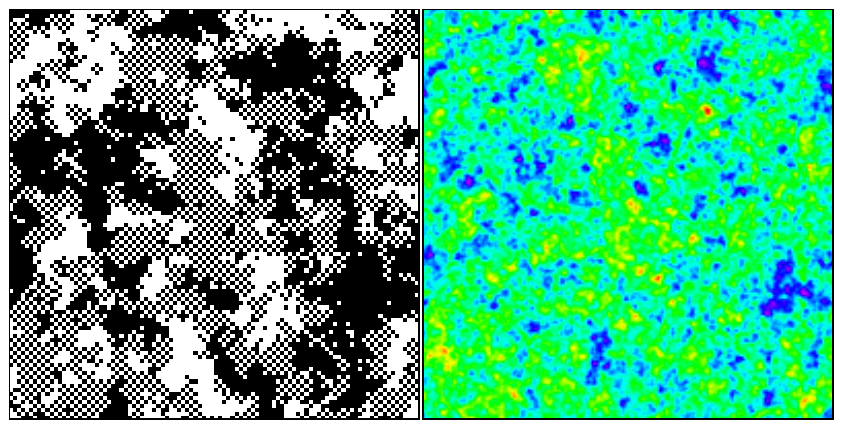}
        {\large $a=0.1$}
        \includegraphics{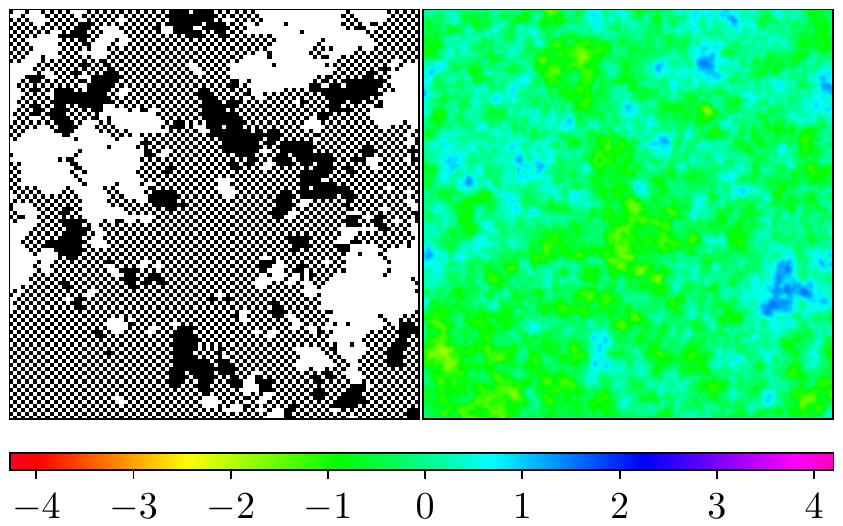}
\caption{(color online)  The right hand side of the figure shows one
  realization of the Gaussian random field. This is a function of continuous
 position but here we show it discretized with
  a spatial resolution of half a lattice spacing.
  From this continuous Gaussian random field,
the bonds are
extracted, at ``half lattice points''. We show the Gaussian field 
for three different correlation exponents. The corresponding GSs
are on the left. The correlations exponents are given by $a\longrightarrow
\infty$ (top), $a=1$ (center) and $a=0.1$ (bottom). Black points denote
$s_{\bm{m}}=-1$ and white points $s_{\bm{m}}=1$. In ferromagnetic order two
neighboring spins have same sign and in antiferromagnetic order the sign
alternates. The system size is $L=100$.}
    \label{fig:gs_examples}
    \end{center}
\end{figure}

To study the spin glass at nonzero temperature we create domain wall (DW)
excitations in the system. This was done by computing GSs under periodic and
antiperiodic BCs also referred to as P-AP \cite{BrayMooreCarter2002}. It works
as follows. First, a spin glass with periodic BCs in one direction and free
BCs in the other direction is generated with quenched disorder,
$\bm{J}^{\text{(p)}}$. Then, its GS configuration,
$\bm{s}_{\text{gs}}^{(\text{p})}$, is computed. Next, the periodic BCs are
replaced by antiperiodic BCs by reversing the sign of one column of bonds
parallel to the direction of periodicity, which leads to
$\bm{J}^{\text{(ap)}}$. Afterwards, the new GS configuration,
$\bm{s}_{\text{gs}}^{(\text{ap})}$, is calculated. The change of the BCs
imposes a DW of minimal energy between the two spin configurations
$\bm{s}_{\text{gs}}^{(\text{p})}$ and $\bm{s}_{\text{gs}}^{(\text{ap})}$. The
energy of the DW is given by \begin{align} \Delta
E=H_{\bm{J}^{(\text{ap})}}\left(\bm{s}_{\text{gs}}^{(\text{ap})}\right)-H_{\bm{J}^{(\text{p})}}\left(\bm{s}_{\text{gs}}^{(\text{p})}\right).
\end{align}

The geometrical structure of a DW can be characterized by the number
$\mathcal{D}_L$ of
bonds which are included in the surface. To avoid
including
those bonds which are a direct result of the different BCs, the surface
is defined to consist of those bonds which fulfill
$J_{\bm{m},\bm{n}}^{(\text{p})} s_{\bm{m}}^{(\text{p})} s_{\bm{n}}^{(\text{p})} J_{\bm{m},\bm{n}}^{(\text{ap})}s_{\bm{m}}^{(\text{ap})} s_{\bm{n}}^{(\text{ap})}<0$,
where $\bm{m},\bm{n}$ runs over all unordered pairs of nearest neighbor lattice sites \cite{KhoshbakhtWeigel2018}.

\section{Results}

We have obtained exact ground states for systems with correlation exponents
in the range $a\in[10^{-3},\infty]$, where $a=\infty$ corresponds to
independently sampled bonds. Since the GS calculation requires
  only polynomial time as a function of the system size, we were
  able to study sizes up to a large value of $N\equiv L^2=724^2$ for
  each value of $a$.
  For each value of $a$ and $L$, we performed an average over many realizations
  of the disorder, ranging from $10^6$ realizations for the smallest sizes,
  10000 realizations for $L=512$, to 2000 realizations
  for the largest system size.

Figure \ref{fig:gs_examples} provides a first impression how the bond
correlation impacts the ordering of the GS. It is apparent that
for strong correlations there are large areas where the spins are either in
ferromagnetic or antiferromagnetic order. This is related to there being large
areas where, due to the correlation, the bonds have identical sign.

\subsection{Domain Wall Energy}

\begin{figure}
    \begin{center}
        \includegraphics{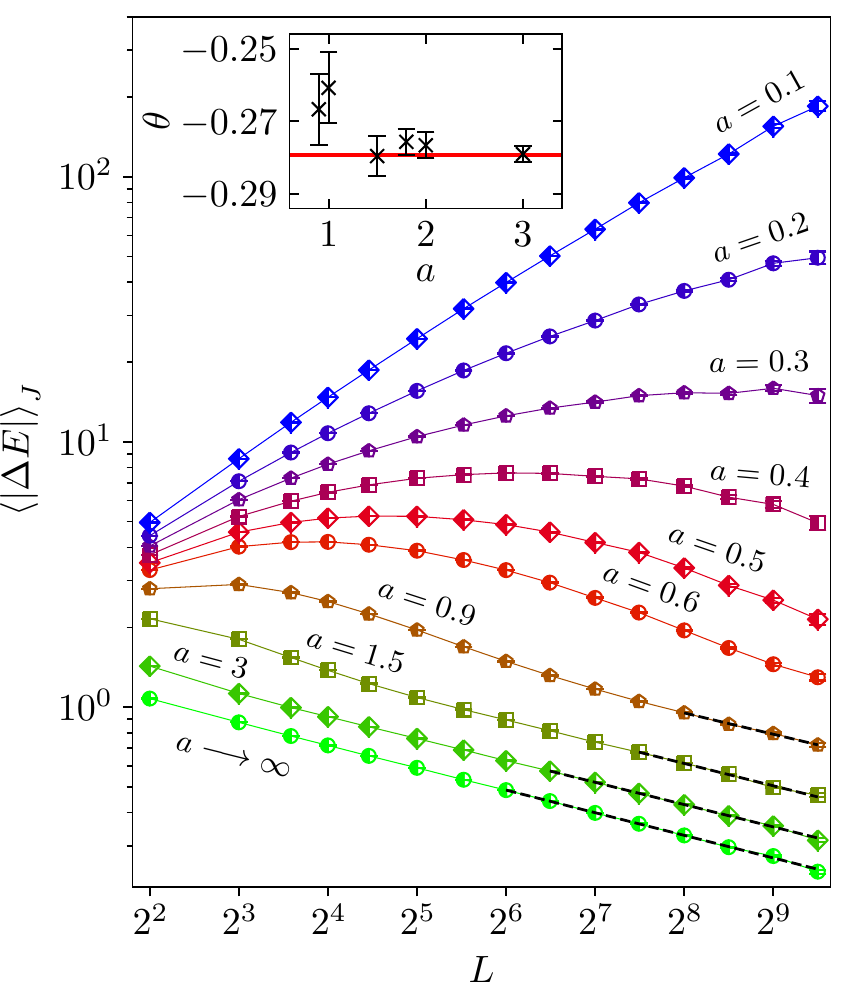}
        \caption{(color online) Scaling of the average of the absolute value of the DW energy,
	$\left\langle \vert \Delta E \vert \right\rangle_J$,
          as a function of system size $L$ for different values of $a$.
          The full lines are guides to the eyes only. The broken lines are fits of type
	  $\left\langle \vert \Delta E \vert \right\rangle_J=A_\theta L^\theta$. The inset shows the values of $\theta$ which were obtained by fits for values of $a\geq 0.9$. The red line marks the value of the stiffness exponent for $a\longrightarrow \infty$ according to \cite{KhoshbakhtWeigel2018}.}
    \label{fig:DwAbsEnergyScaling}
    \end{center}
\end{figure}

Next, we look at the influence of bond correlation on the properties of the previously discussed DW excitations. The absolute value of the DW energy is proportional to the coupling strength between block spins in the zero-temperature limit \cite{BrayMoore1987}. A stable order is possible if the average absolute value of the DW energy increases with the system size. In the uncorrelated case, $a\longrightarrow \infty$, one obtains a power law behavior \cite{BrayMoore1987,HartmannYoung2001,aspect-ratio2002,KhoshbakhtWeigel2018},
\begin{align}
    \langle\vert  \Delta E \vert\rangle_J \sim L^{\theta},
    \label{eq:AbsEnergyScaling}
\end{align}
where $\theta$  is the stiffness exponent with its current best estimate
$\theta=-0.2793(3)$ \cite{KhoshbakhtWeigel2018}. Since $\theta<0$ there is no stable spin-glass phase for temperatures larger than zero. Figure \ref{fig:DwAbsEnergyScaling} shows the impact of
bond correlation on the scaling of $\langle\vert  \Delta E \vert\rangle_J$.
For $a\geq 0.9$ one can still observe the pure power-law decay of the
uncorrelated model on sufficiently long length scales. The inset demonstrates
that, in this region, the stiffness exponent stays constant at a value equal to that of the
uncorrelated model. For values of $a\leq 0.9$ the average $\langle \vert
\Delta E \vert \rangle_J$ initially grows with system size, but starts to
decrease for larger sizes, i.e. the curves  exhibit a peak. The system size
at the peak,
$L^*$, shifts to larger system sizes on decreasing the
correlation exponent $a$.  This will be analyzed below.

\begin{figure}
    \begin{center}
        \includegraphics{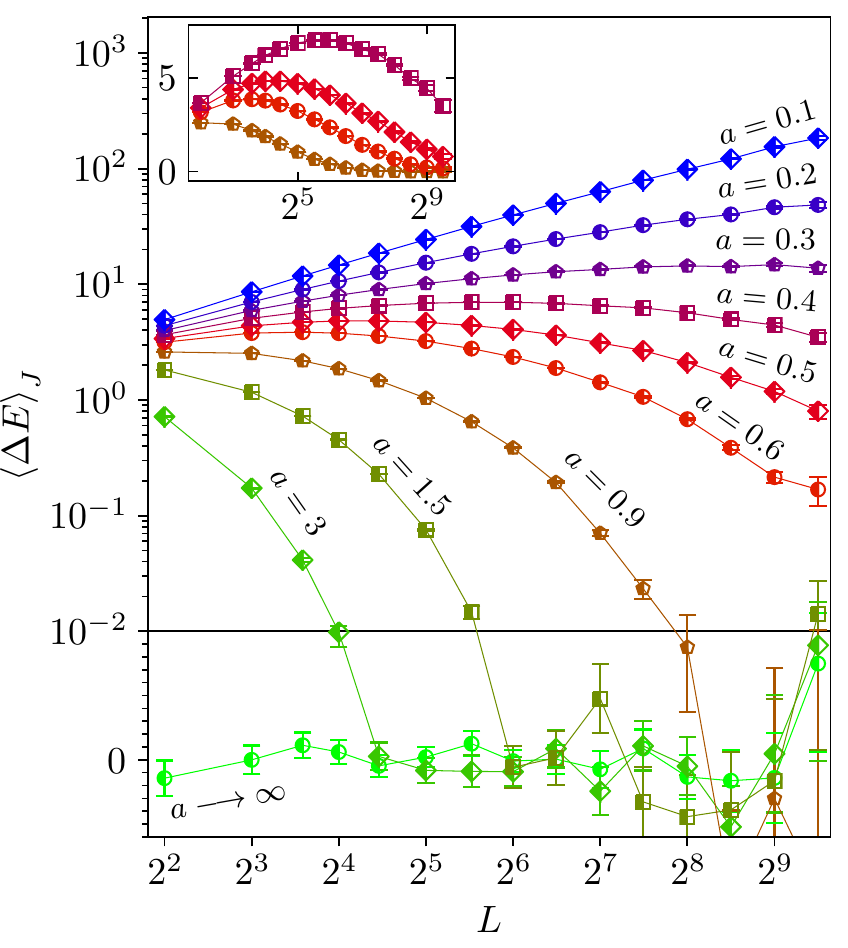}
        \caption{(color online) A log-log plot of the average DW energy
          $\left \langle \Delta E \right \rangle_J$ as a function of the
          system size $L$ for some values of
          correlation exponent $a$.
          The inset shows the same data with linear energy scale
          for $a=0.4$, 0.5, 0.6 and 0.9 to highlight the peak structure.
 The lines are guides to the eyes only.}
    \label{fig:DwEnergyScaling}
    \end{center}
\end{figure}

First we show in figure \ref{fig:DwEnergyScaling} the results for the average DW energy
$\langle \Delta E \rangle_J$, which behaves in a somewhat similar manner as
the average of the absolute value. 

If spin-glass order existed in this model for some value of $a$ one would
observe an increase $\langle \vert \Delta E \vert \rangle_J$
in the limit $L\to\infty$, while at the same time
$\langle \Delta E \rangle_J$  would remain at, or converge to,
zero as a function of $L$. The latter indicates the absence of 
ferromagnetic order, but not necessarily the absence of spin glass order
because for a spin glass the
change of the boundary conditions from periodic to antiperiodic
is symmetric, i.e., could either increase or decrease the GS energy, so the
average would be zero even in the case of spin glass order.
An increase of both $\langle \vert \Delta E \vert \rangle_J$ and $\langle
\Delta E \rangle_J$ for small values of $L$ and $a$ corresponds to a
ferro/antiferromagnetic ordering on local length scales, which is visible in
Fig.\ \ref{fig:gs_examples} and will be discussed more below. Whether there
is a true ordered phase for very small values of $a$ will be discussed next.

\begin{figure}
    \begin{center}
        \includegraphics{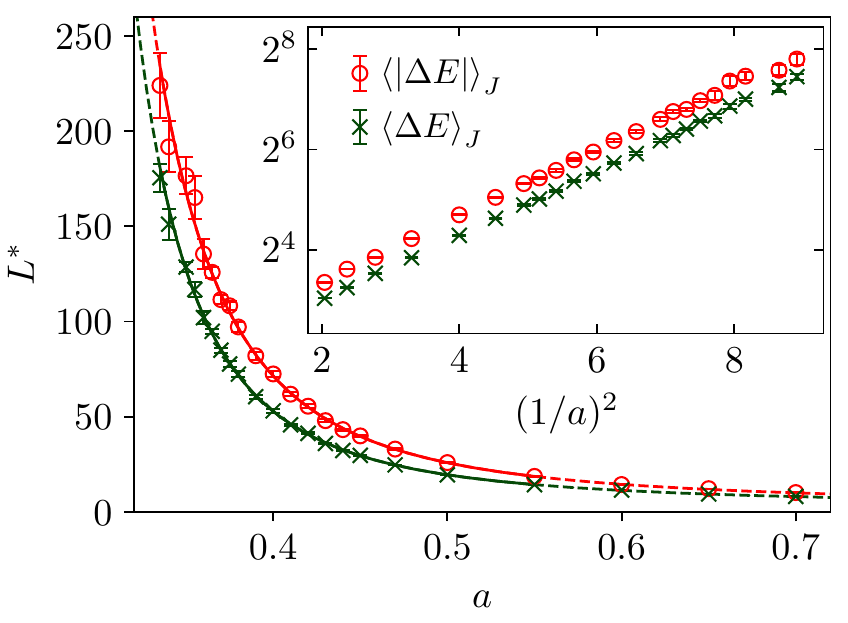}
        \caption{(color online)
          The system size where the peak of
          $\langle \vert \Delta E\vert \rangle_J$
          and $\langle\Delta E \rangle_J$ occurs,
      denoted by $L^*$, as a function of $a$. The lines are fits
      according to Eq. \eqref{eq:peak_fit} in the range $a\le 0.55$,
      see text for details.
      The broken lines are
      extrapolations of the fits. The inset shows $L^*$ on a
      logarithmic scale as a function of $1/a^2$ exhibiting a
      straight-line behavior, and thus
      confirming the behavior obtained from the fit.}
    \label{fig:LocationOfPeak}
    \end{center}
\end{figure}

To track how the length scale of local order changes as a function of the
correlation strength we measure the size at the peak, $L^*$, for both the
domain-wall energy and the absolute value, as a function of the correlation
exponent $a$. Numerically, $L^*$ was computed by fitting a parabola in the
vicinity of the peaks of $\langle \vert \Delta E \vert \rangle_J$ and $\langle
\Delta E \rangle_J$. As the fit in figure \ref{fig:LocationOfPeak}
demonstrates, the data for $L^*(a)$ is well described by an exponential
function
\begin{align}
L^*(a)= A_{L} \exp\left\{ b_L\left(a-a_{\text{crit}}\right)^{-c_{L}} \right\}.
     \label{eq:peak_fit}
\end{align}
A non-zero value of $a_{\text{crit}}$ would indicate that
an ordered phase exists for $a<a_{\text{crit}}$.
A true spin-glass phase would be possible
if $a_{\text crit}$ for $\langle  \Delta E \rangle_J$
is smaller than $a_{\text crit}$ for 
$\langle \vert \Delta E\vert \rangle_J$. We obtained 
values of $a_{\text{crit}}=0.13(0.10)$ for $\langle  \Delta E \rangle_J$ and
$0.10(0.17)$ for $\langle \vert \Delta E\vert \rangle_J$ with quality of the
fit $Q=0.87$ and $Q=0.58$, respectively. Thus, a zero value for the critical
correlation exponent parameter $a_c$ seems likely.

To consider this further, we set $a_{\text{crit}}$ to zero and obtain the
values of the other fit parameters, which here are $A_L=3.63(22)$,
$b_L=0.39(4)$, $c_L=2.10(6)$ ($Q=0.86$) for $\langle  \Delta E \rangle_J$ and
$A_L=3.5(4)$, $b_L=0.55(6)$, $c_L=1.85(8)$ ($Q=0.63$) for $\langle \vert
\Delta E\vert \rangle_J$. Since the qualities of the fits remain almost
identical in comparison to $a_{\text{crit}}\neq0$ the data is considered to be
consistent with $a_{\text{crit}}=0$, which would imply that there is no global
order when $a>0$, either ferromagnetic or spin glass. The inset shows that
the behavior of $L^*(a)$ is also compatible with $c_L=2$. Fits of this type,
with $a_{\text{crit}}=0$ and $c_L=2$ fixed, have quality of the fit larger
than $0.4$, which is reasonable. Note that an exponential dependence of the
``breakup'' length  scale of ferromagnetic order as a function of disorder
strength was also found in the two-dimensional random field Ising model, both at
low temperatures \cite{PytteFernandez1985} and in the GS
\cite{SeppalaEtAl1998}.

\subsection{Domain Wall Surface}

\begin{figure}
    \begin{center}
        \includegraphics{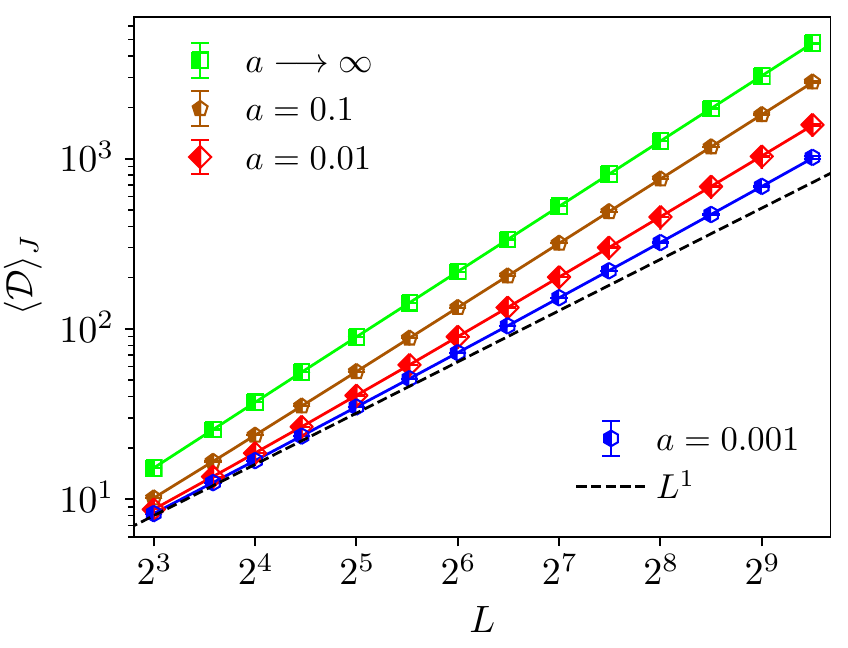}
    \caption{(color online) Scaling of the DW surface area $\langle \mathcal{D}\rangle_J$ for different values of $a$. The broken black line shows the scaling of the DW surface in case of a ferro/antiferromagnet. The full lines follow fits according to Eq. \eqref{eq:surface_scaling_fit} with $L_{\text{min}}^{(\text{fit})}=8$.}
    \label{fig:surface_scaling}
    \end{center}
\end{figure}

The behavior of DW surfaces is regarded as one of the essential parameters
which describe the properties of random systems
 \cite{BrayMoore1987choatic,NewmanStein2007}.
 DWs separate spins in GS and reversed GS. Their surface
 is defined as those bonds which belong to the DW, and we denote
 the surface size by $\mathcal{D}$.

 In the uncorrelated case, $a\longrightarrow \infty$,
 the average DW surface size exhibits a power law
 \cite{kawashima2000,fract_dim_DW2007},
 \begin{align}
\langle \mathcal{D} \rangle_J \sim L^{d_s},
\label{eq:surface_scaling}
\end{align}
where $d_s$  is the fractal surface dimension, for
which the best numerical estimate at present is
$d_s=1.27319(9)$ \cite{KhoshbakhtWeigel2018}. When $a=0$ the system is a
ferro/antiferromagnet and $\langle \mathcal{D} \rangle_J=L$, implying that
$d_s=1$, i.e., the surface is not fractal here.  Figure
\ref{fig:surface_scaling} shows the scaling of the DW surface size for
different values of $a$. In general, it can be seen that the correlation
decreases the number of bonds in the DW surface. Even for $a=0.001$, the data on
this log-log plot shows
a visible a visible deviation from the linear behavior which occurs for $a$
strictly zero.

\begin{figure}
    \begin{center}
        \includegraphics{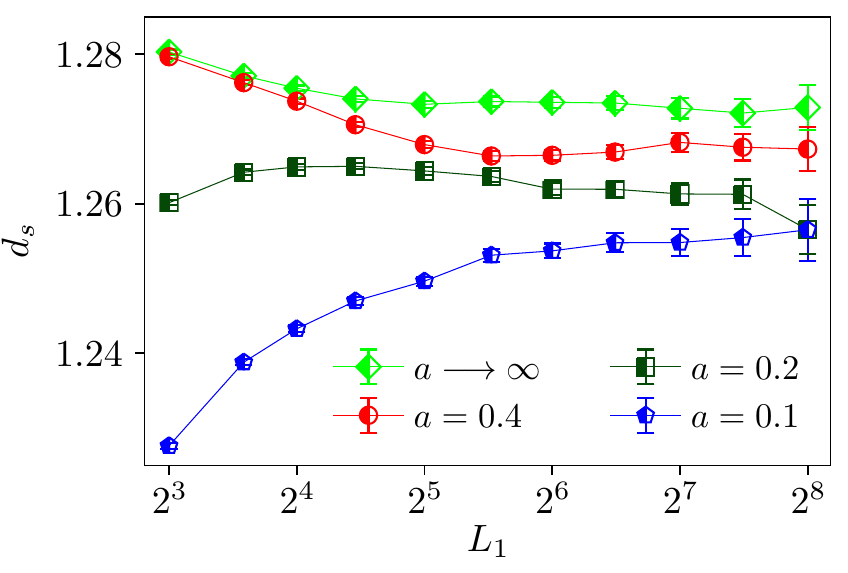}
        \caption{(color online) The fractal surface dimension, $d_s$, of the DW
          surface area as a function of the smallest system size $L_1$ of
          a fit window.}
    \label{fig:ds_fit_window}
    \end{center}
\end{figure}

To characterize this behavior we fitted pure power-laws to the data.
For this purpose, we did not do a full fit to all data points, but used
the sliding-window approach instead. Here,
four values of $\langle \mathcal{D}\rangle_J$ which are adjacent in terms
of system size
were grouped together in one fit window, respectively.
The independent variables of such a fit window were given by $(L_1,L_2,L_3,L_4)$ with $L_i<L_{i+1}$, $i=1,...,4$. The dependent variables corresponded to the data, i.e. $\langle \mathcal{D}_{L_i}\rangle_J$. The smallest independent variable of each fit window is denoted as $L_1$. 

\begin{figure}
    \begin{center}
        \includegraphics{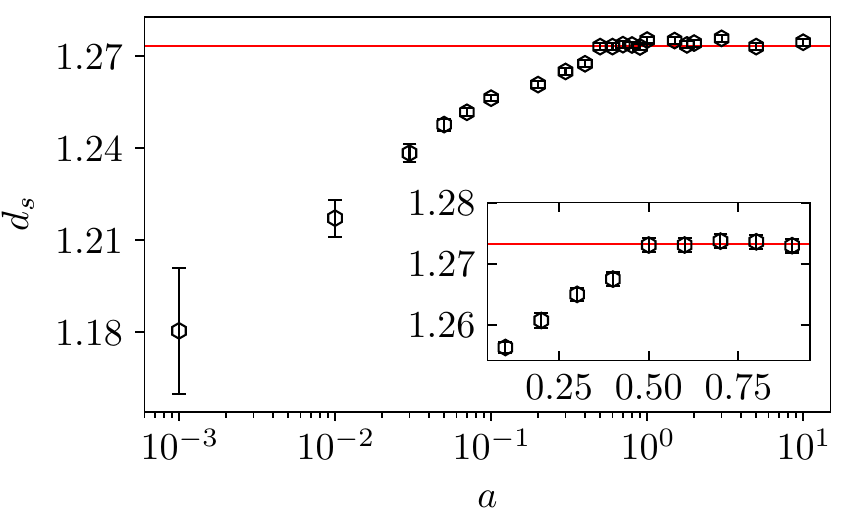}
	\caption{(color online)  The fractal surface dimension as a function
	of $a$. The values for $a\geq0.2$ were obtained by pure power law
	fits, i.e. $\langle \mathcal{D} \rangle_J=A_{\mathcal{D}} L^{d_s}$
	with smallest system size used in the fits
	$L_{\text{min}}^{(\text{fit})}=128$. The values for $a\leq 0.1$ were
	obtained from fits to Eq. \eqref{eq:surface_scaling_fit}, which
	includes the leading correction to scaling, with
	$L_{\text{min}}^{(\text{fit})}=8$. The red line marks the value of
	$d_s$ for $a\longrightarrow \infty $ according to
	\cite{KhoshbakhtWeigel2018}.}
    \label{fig:fractal_surface_dimension}
    \end{center}
\end{figure}

For each window, we fitted the
power law to the data resulting in a value of $d_s$.
The dependence of $d_s$  as a function of $L_1$ for different values of $a$ can be found in figure \ref{fig:ds_fit_window}. In the uncorrelated case $d_s$ decreases as a function of $L_1$, whereas for strong correlations $d_s$ increases.
In any case, for system sizes $L_1< 32$ and small values $a\le 1$
we observed some notable dependence
of $d_s$ on the system size. This motivated us to include
corrections to scaling the power law behavior in
Eq. \eqref{eq:surface_scaling} by considering
 \cite{KhoshbakhtWeigel2018,MooreHartmann2003Corrections}
\begin{align}
    \left\langle \mathcal{D}  \right\rangle_J=A_\mathcal{D} L^{d_s}(1+B_\mathcal{D}L^{-\omega_s})~.
    \label{eq:surface_scaling_fit}
\end{align}
By using Eq. \eqref{eq:surface_scaling_fit} the quality of the fit is larger than $0.79$ for all studied values of $a \leq0.1$ with smallest linear system size of the fit $L_{\text{min}}^{(\text{fit})}=8$. For larger values of $a$,
the pure linear fit was always fine, given our statistical accuracy.
Figure \ref{fig:fractal_surface_dimension} shows the resulting fractal surface dimension for all considered values of $a$. At $a\approx 0.4$ the fractal surface dimension starts to decline from $d_s=1.27319(9)$ \cite{KhoshbakhtWeigel2018}, in the uncorrelated case, to smaller values. This is also visible
in Fig.~\ref{fig:ds_fit_window}. Thus, it appears that for small
values of $a$ the fractal structure of the cluster changes, although
there is no phase transition.  Note that, in order to extract $d_s(a)$,
we also performed fits
of the form $d_s(L_1;a)=d_s(a)+\kappa_d L_1^{\gamma_d}$
(not shown; $\kappa_d$ and $\gamma_d$ also depend on $a$)
to the sliding window fractal dimensions shown
in Fig.~\ref{fig:ds_fit_window}. The behavior of this
extrapolated fractal dimension also exhibits the same notable decrease
of $d_s$ for $a \le 0.4$.
Of course, it can not be ruled out that on sufficiently large length scales, i.e.,
for much
larger sizes than are currently accessible, the fractal dimension of the
uncorrelated model would be recovered again and thus
$d_s=1.27319(9)$ \cite{KhoshbakhtWeigel2018} for all values $a>0$.

\subsection{Spin Correlations in the Ground State}
In this section the effects of the bond correlations on spin
correlations in the GS will be discussed. Note that at zero temperature
there is no thermal disorder and, in our study which uses bonds with a
continuous distribution,
the ground state is non-degenerate apart from overall spin inversion.

The two-point spin correlation is given by $\langle  s_{\bm{m}}^{(\text{gs})}
s_{\bm{m}+\bm{r}}^{(\text{gs})} \rangle_J$, where $s_{\bm{m}}^{(\text{gs})}$
denotes a spin in the GS configuration. In the uncorrelated model $\langle
s_{\bm{m}}^{(\text{gs})} s_{\bm{m}+\bm{r}}^{(\text{gs})} \rangle_J=0$ if
$\bm{r}\neq 0$. The correlated bonds induce a local ferro/antiferromagnetic
order into the GS. When $a=0$ the system is a ferro/antiferromagnet and the
order is global. In the ferromagnetic case $s_{\bm{m}}^{(\text{gs})}
s_{\bm{m}+\bm{r}}^{(\text{gs})}=1$ and in the antiferromagnetic case
$s_{\bm{m}}^{(\text{gs})} s_{\bm{m}+\bm{r}}^{(\text{gs})}$ alternates between
plus and minus one. Therefore, the GS spin correlation can be estimated by
\begin{align}
G_{\text{gs}}(\bm{r})&= \frac{1}{\vert \Lambda'(\bm{r}) \vert}\sum_{\bm{m} \in\Lambda'(\bm{r})}  \left(\frac{ \hat{\sigma}(\bm{r})+1}{2} \right)~\langle s_{\bm{m}}^{(\text{gs})} s_{\bm{m}+\bm{r}}^{(\text{gs})}  \rangle_J~,
\label{eq:gs_correlation} \\
\hat{\sigma}(\bm{r})=&\sigma(r_1)\sigma(r_2)~\text{with}~\sigma(r_i)=
\begin{cases}
    1~~&\text{if}~r_i~\text{is even} \\
    -1~~&\text{if}~r_i~\text{is uneven}
\end{cases} \nonumber
\end{align}
and $\bm{r}=(r_1,r_2) \in \mathbb{Z}^2$. $\Lambda'(\bm{r}) \subset \Lambda$,
similar to Eq.~(\ref{eq:bond_corr_estimator}), contains those
sites $\bm{m}$ for which $\bm{m}+\bm{r}$ is on the lattice, given the free
BC in one direction.
Note that this definition means that the correlation
is measured for each site
$\bm{m} \in\Lambda'$ on one of the two sublattices of a checkerboard partition
of the square lattice, such that the correlation is insensitive to whether the
order is ferromagnetic or antiferromagnetic.

\begin{figure}
    \begin{center}
        \includegraphics{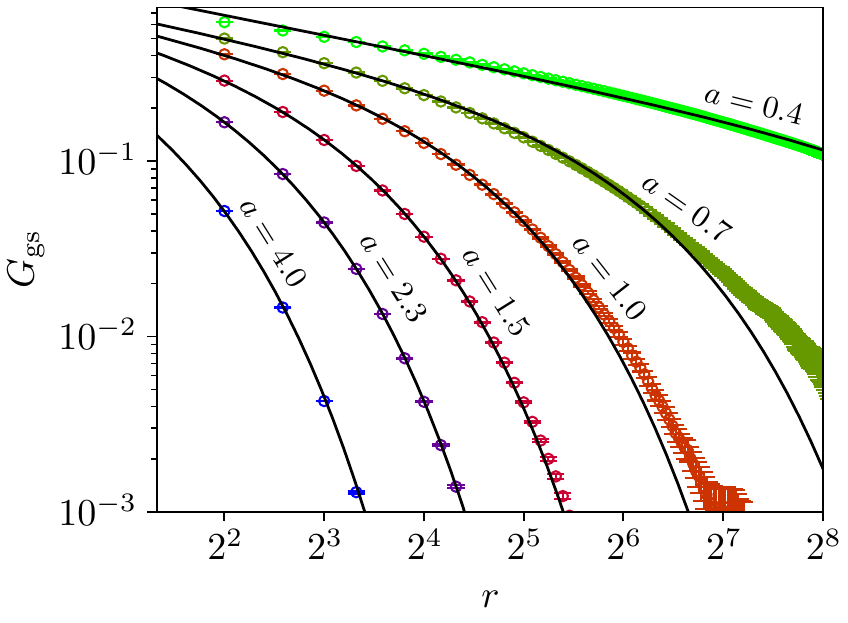}
        \caption{(color online) Spin correlation of the GS for different values of $a$. The correlation was computed by utilizing the estimator of Eq. \eqref{eq:gs_correlation} along the main axes.
          The black lines are fits according to Eq. \eqref{eq:gs_correlation_function}
          when using a multifit, i.e., fitting two exponents $\upsilon$
          and $\varphi$ and many values of $\xi_{\text{gs}}(a)$ simultaneously
          to the correlation obtained for all values of $a$.
        }
    \label{fig:gs_correlation}
    \end{center}
\end{figure}

In figure \ref{fig:gs_correlation} one can see the GS correlation for different values of $a$.
The data is well described by a scaling form of type
\begin{align}
G_{\text{gs}}(r)\sim \frac{1}{r^{\upsilon}} \exp\left\{- \left( \frac{r}{\xi_{\text{gs}}} \right)^\varphi \right\}\,.
    \label{eq:gs_correlation_function}
\end{align}
From the perspective of ordering we are especially interested in the
correlation length, $\xi_{\text{gs}}$, that provides the distance over which
spins are notably correlated. The straightforward method to extract
$\xi_{\text{gs}}$ is by fitting the function of Eq.
\eqref{eq:gs_correlation_function} to the data. The problem with such an
approach is that neither $\upsilon$ nor $\varphi$ are known. Also finite-size
corrections reduce the match between scaling form and actual data.  Thus, we
used different approaches to obtain $\xi_{\text gs}$.

First,
we perform a separate \emph{fit} for each value of $a$  down to
small correlations where the error bars start to exceed one quarter of the
correlation value. We observed that
for $a\ge 2$, the values of the exponents $\upsilon$ and $\varphi$
did not change much, while for smaller values of $a$ the exponents
were a bit smaller.
Therefore, we fixed the exponents to the (averaged) values seen for $a\ge 2$
and fitted only with respect to $\xi_{\text{gs}}$.

\begin{figure}
    \begin{center}
        \includegraphics{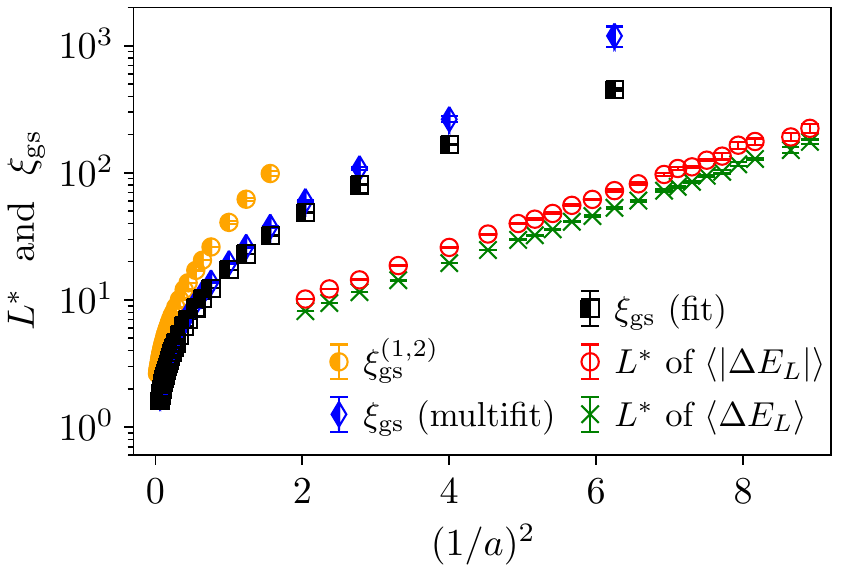}
        \caption{(color online) The correlation length of the GS,
          $\xi_{\text{gs}}^{(1,2)}$, as a function of $1/a^2$,
          obtained by three different approaches. For comparison the
        length scales $L^\star$ of the maxima are shown again.}
    \label{fig:GsCorrelationLengthAndLocationOfMaxima}
    \end{center}
\end{figure}

Second, we also performed
a {\em multifit}, i.e., we fitted the correlation function simultaneously
\cite{practical_guide2015}  with one value for $\upsilon$, one value for
$\varphi$ and values of the correlation length at many values of $a$.
The results
of this multifit are also shown in Fig.~\ref{fig:gs_correlation}.
The obtained values for $\xi_{\text{gs}}$ for these two fitting
approaches are shown in
Fig.~\ref{fig:GsCorrelationLengthAndLocationOfMaxima}. As can be seen,
the results from fixing the values of the exponents and from using the
multifit approach do not differ much.

Before we discuss the behavior of $\xi_{\text{gs}}(a)$, we
describe the third approach we have used
to estimate the correlation length. Here, we used
the \emph{integral estimator} that was introduced in
Ref.~\cite{BellettiEtAl2008Nonequilibrium}. It presupposes that
for $r\leq \xi_{\text{gs}}$ the correlation function is dominated
by a power law of type  $r^{-\upsilon}$, whereas for $r>\xi_{\text{gs}}$
the correlation is negligible. As a consequence, the integral
\begin{align}
    I_k=\int_0^\infty \text{d}r\, r^k G_{\text{gs}}(r)
    \label{eq:weighted_correlation_intergral}
\end{align}
is given by $I_k \propto (\xi_{\text{gs}})^{{k+1}-\upsilon}$ and thus
\begin{align}
    \xi_{\text{gs}}^{(k,k+1)}:=\frac{I_{k+1}}{I_{k}}\propto \xi_{\text{gs}} \,.
\end{align}
This result would be exact and independent of $k$ if the correlation
actually was only a power law. For real correlations,
the value of $k$ dictates which part of the correlation function contributes most to the integral. Because such a scaling approach is only valid when $r$ is much larger than the lattice constant a high value of $k$ reduces the systematic error of the method. On the other hand, large values of $k$ increase the statistical error of $I_k$. Following the recommendation of \cite{Belletti2009AnIndepthView} we used $\xi_{\text{gs}}^{(1,2)}$ as a compromise. Note that
since the statistical error of the measured correlation grows with
distance $r$, one usually defines a cutoff distance up to which the
data is directly used for the integral.
Similar to \cite{Belletti2009AnIndepthView} we specified this cutoff
distance as the value of $r$ where $G_{\text{gs}}$ is smaller than three times
its error. For values of $r$ larger than this cutoff distance we computed the
integral up to the maximal length of $L$ from fits to Eq.
\eqref{eq:gs_correlation_function}. The start value of these fits were set to
$r_{\text{min}}^{(\text{fit})} \geq 2$. Because it was observed that the
correlation decays slightly differently along the directions with free and
periodic BCs, the computations of the GS were also done with an independent set of
simulations for full free BCs.

Next, the correlation length $\xi_{\text{gs}}^{(1,2)}$ was
extracted from the average of the GS correlation, $G_{\text{gs}}$,
along the main axes. The statistical error of
$\xi_{\text{gs}}^{(1,2)}$ was estimated by bootstrapping
\cite{Young2015Everything,practical_guide2015} and the integrals
according to Eq. \eqref{eq:weighted_correlation_intergral} were
computed by utilizing the midpoint integration rule. For small values
of $a$, the contribution to $I_k$ from the integral beyond the cutoff
gets increasingly large. For instance, when $a=1.15$ this contribution
made up approximately $4\%$ of the value of $I_2$. Hence, to estimate
the total error, we added an extra systematic error to the
statistical error. This was done by analyzing the value of
$\xi_{\text{gs}}^{(1,2)}$ for two other choices of the cutoff distance,
i.e., being the distance where the error of the
correlation function is two or four times larger than its
estimate, respectively.
The maximal deviation of these two values from the standard definition, which
uses a magnitude of three error bars to define  cutoff,
was set to be the systematic
error. Furthermore, because the statistical error of $I_k$ grows large
for small values of the correlation exponent $a$
and the system has to be sufficiently large to neglect
boundary effects, values of $a<0.8$ were not considered for this approach.

Figure \ref{fig:GsCorrelationLengthAndLocationOfMaxima}
shows $\xi_{\text{gs}}^{(1,2)}$ as a function of the correlation exponent.
In contrast to the data for the length scales $L^\star$,
the plot shows that the behavior of none of the three correlation lengths that
we have defined,
$\xi_{\text{gs}}\ \text{(fit)},\  \xi_{\text{gs}}\ \text{(multifit)}$ and
$\xi_{\text{gs}}^{(1,2)}$,
is close to 
exponential. Nonetheless, for $a\to 0$ the correlation lengths
may converge to this behavior.
We found that the data for these three definitions of the correlation length could be
well fit to a function of the type
\begin{equation}
\xi_{\text{gs}}=A_{\xi} (a-a_{\text{crit}})^{-d_{\xi}}
\exp \left\{ b_\xi (a-a_{\text{crit}})^{-c_{\xi}} \right \}.
\label{eq:gs_correlation_length_fit}
\end{equation}
The exponential in Eq. \eqref{eq:gs_correlation_length_fit}
is consistent with the previous results for the length scale $L^*$ and
will dominate for $a\to 0$. The power-law part is chosen to describe
the behavior for large values of $a$.

We have fitted the data to this function. For example, 
for the data obtained from the multifit we again obtained a
small value of $a_{\text{crit}}=0.11(6)$.
Thus once again we fixed $a_{\text{crit}}\equiv 0$ in order to determine
the values of the remaining fit parameters, obtaining $A_{\xi}=4.0(3)$,
$d_{\xi}=0.81(3)$, $b_{\xi}=1.58(8)$ and $c_{\xi}=1.19(4)$ with a good
quality of the fit. Similar values, in particular for $c_{\xi}$,
are found for the other two definitions of $\xi_\text{gs}$ that we used.
This means that for large values of the correlation exponent
$a$ the behavior of the correlation length as function of $a$
seems to be better described by a power law. Also, the behavior of the
peak lengths and of the ferromagnetic correlation lengths differ a lot,
but it is still possible that for very small values of $a$, an exponential
dependence of $\xi_{\text{gs}}$ on $1/a^2$ would be recovered. Unfortunately, to
investigate this issue
much larger system sizes would have to be treated, well beyond 
current numerical capabilities.

\section{Discussion}
The standard
two-dimensional Ising spins glass does not exhibit a finite-temperature spin glass phase in contrast to the three or higher dimensional cases \cite{YoungStinchcombe1976,YoungSouthern1977,HartmannYoung2001}. This work deals with the question how long-range correlated bonds influence this characteristic. Therefore, the ordering behaviour of the two-dimensional Ising spin glass with spatially long-range correlated bonds is studied in the zero-temperature limit. The bonds are drawn from a standard normal distribution with a two-point correlation for bond distance $r$ that decays as $(r^2+1)^{-a/2}$, $a \geq 0$. In the borderline case, when $a=0$, the system is either a ferromagnet or antiferromagnet, depending on the bond realization. For $a\to \infty$ the uncorrelated EA model is recovered.

For $0<a<\infty$ we observed that the correlation has local effects
on the zero-temperature ordering behaviour. The correlation locally effects
the average value of the bonds as well as their standard deviation for
each individual  realization of the disorder. These parameters are decisive to distinguish between a spin glass or ferro/antiferromagnet in case of the uncorrelated model \cite{SherringtonSouthern1975,GarelMonthus2014}. In correspondence to that, the spin correlation of the GS reveals how the correlation induces a local ferro/antiferromagnetic order into the GS. This is reflected by
a growing correlation length $\xi_{\text{gs}}(a)$ when decreasing $a$.

Complementary results to the direct study of the GS spin configurations were obtained by
investigating DW excitations. The average of the absolute value of the DW
energy can be interpreted as the coupling strength between block spins at zero
temperature \cite{BrayMoore1987}. We found, that for strong bond correlation, the
average of
absolute value of the DW energy initially increases as a function of the system size up to a 
peak, and then decreases. Since we made the same observation for the actual
DW energy it shows that the increase of the absolute value of the DW energy is
a consequence of local ferro/antiferromagnetic order of the system in GS. The
system size where the peak occurs, $L^*$, is interpreted as the length scale
of local order. For small values of the correlation exponent $a$, both $L^*(a)$ and the
correlation length of the GS, $\xi_{\text{gs}}(a)$, can be described by
an exponential divergence.
Interestingly, a similar exponential length scale was also found in the
two-dimensional random field Ising model by GS computations
\cite{SeppalaEtAl1998} and at low temperatures \cite{PytteFernandez1985}. In
these studies the length scale of ferromagnetic order was examined as a
function of the standard deviation of the random magnetic field.

For the two-dimensional Ising spin glass, the distribution of the absolute value of the domain wall energy is ``universal'' with respect to the initial bond distribution. This means for any continuous, symmetric bond distribution with sufficiently small mean and finite higher moments, the absolute value of the domain wall energy should approach the same scaling function \cite{Mcmillan1984,BrayMoore1987,AmorusoEtAl2003Scalings}. Thus, we expect the same kind of universality for our model.

The stiffness exponent $\theta$ describes the scaling of the width of this distribution,
and is related to the critical exponent $\nu$ describing the 
divergence of the correlation length as $T \to 0$ by
$\nu=-1/\theta$ \cite{FernandezEtAl2016,BrayMoore1987}. At this zero
temperature transition $\nu$ is the only independent exponent. Therefore, any
bond correlation which
leaves the stiffness exponent unchanged does not influence the universality of
the model. From the data of $L^*$ it is expected that there is no global
ordered phase for $a>0$. This implies that the stiffness exponent is negative
for $a>0$. Furthermore, for values of $a$ in the range $a\geq 0.9$ the stiffness exponent
stays equal to the uncorrelated case. Due to the limited range of
studied length scales it was not possible for us to verify this for values
of $a$ less than $0.9$.

In addition to the DW energy, another important parameter to
describe the low-temperature behaviour of the Ising spin glass is the DW
surface area
\cite{BrayMoore1987choatic,NewmanStein2007}. In the uncorrelated model the
domain-wall surface area follows a power law, $\langle \mathcal{D} \rangle_J \sim
L^{d_s}$, where  $d_s=1.27319(9)$ \cite{KhoshbakhtWeigel2018}. Our results are
compatible to this for all considered correlation exponents $a\geq0.5$. At
$a\approx0.4$ the fractal surface dimension starts to decline. For the values
of $a\leq0.1$ the data is well described by a power law with scaling
corrections \cite{MooreHartmann2003Corrections,KhoshbakhtWeigel2018}, i.e.
$\langle \mathcal{D} \rangle_J=A_\mathcal{D}
L^{d_s}(1+B_\mathcal{D}L^{-\omega_s})$.  The decrease in $d_s$ implies
that for strong correlations DWs with shorter lengths are energetic favorable.
In the extreme case when $a=0$ the system is a ferro/antiferromagnet and
thus $d_s=1$. 
Of course, it can not be ruled out that the decline in $d_s$ is local and on
sufficiently long length scales the pure power law with $d_s=1.27319(9)$ \cite{KhoshbakhtWeigel2018} is recovered again for all $a>0$.

In this context it is interesting to note that there exists a
proposed relation which links the stiffness exponent with the fractal surface dimension, namely
$d_s=1+3/(4(3+\theta))$ \cite{AmorusoEtAl2006}. According to highly accurate
numerical results \cite{KhoshbakhtWeigel2018} this equation is probably not
exact. Our results for the fractal surface dimension $d_s=1.27318(29)$
($Q=0.99$) and the stiffness exponent $\theta=-0.2815(13)$ ($Q=0.13$) deviate
by approximately 6.5 standard deviations from the mentioned conjecture, since
$d_s-1-3/(4(3+\theta))=-0.0027(4)$. The latter result was obtained by using
standard error propagation, thus, neglecting correlations between the
estimates of $d_s$ and $\theta$ which exist because both values were obtained
from the same data set. Nonetheless, our results also support the conclusion that the
proposed scaling relation is not exact.

In conclusion, it is observed that correlation among the bonds has strong
effect on the ordering on local length scales, inducing
ferro/antiferromagnetic domains into the GS. The length scale of local
ferro/antiferromagnetic order diverges exponentially when the correlation
exponent approaches zero. The fractal surface dimension decreases for strong
correlations on the studied length scales. No signature of a spin-glass phase at
finite temperature is observed.

\acknowledgements
        The simulations were performed at the HPC cluster CARL, 
        located at the University of Oldenburg (Germany) and funded by the DFG
        through its Major Research Instrumentation Programme
        (INST 184/157-1 FUGG) and the Ministry of
        Science and Culture (MWK) of the Lower Saxony State.

\section*{ }
\bibliography{corr_sg}
\end{document}